\documentclass[twocolumn,prb,showpacs]{revtex4}
 \usepackage{graphicx}

\begin{document}

 \title{Solitons in polarized double layer quantum Hall systems}

\author{R. Khomeriki}
\altaffiliation[Also at ]{Department of Physics, Tbilisi State University, Tbilisi,
 Georgia}

 \author{M. Abolfath}
 \altaffiliation[Also at ]{Center for Semiconductor Physics in Nanostructures}

 \author{K. Mullen}
 \altaffiliation[Also at ]{Center for Semiconductor Physics in Nanostructures}
 \affiliation{University of Oklahoma, Department of Physics and Astronomy,
 440 W. Brooks, 73019 OK}

 \date{\today}

 \begin{abstract}

  A new manifestation of interlayer coherence in strongly
 polarized double layer quantum Hall systems with total filling factor
 $\nu=1$ in the presence of a small or zero tunneling is theoretically
 predicted.  It is shown that moving (for small tunneling) and spatially
 localized (for zero tunneling) stable pseudospin solitons develop which
 could be interpreted as mobile or static charge-density excitations.
 The possibility of their experimental observation is also discussed.

 \end{abstract}

 \maketitle

 Much study has been devoted to the dynamical properties of
 charge-density excitations in double layer quantum Hall systems with
 total Landau filling factor $\nu=1$.  An effective field-theoretical
 pseudospin model was developed in order to explain the experimentally
 observed spontaneous phase coherence even in the absence of tunneling
 \cite{moon,yang,brey} as well as for nonzero tunneling and in the presence
 of an applied in-plane magnetic field \cite{sheena}.  The main
 prediction \cite{stern,balents,fogler} of this model, the existence of
 the broken symmetry Goldstone mode \cite{fertig}, has been confirmed by
 the observation \cite{spielman} of split off peaks in the tunneling
 conductance in the presence of in-plane magnetic field.

 However, the study of double layer (pseudo) ferromagnets is usually
 restricted to the charge balanced state (see e.g.  \cite{hanna} and
 references therein), although very recently the existence of  a stable
 unbalanced state (the so called canted phase \cite{leo,ramin1}) in the presence
 of in-plane magnetic field was predicted. To date,
 static structures have dominated theoretical and experimental
 investigations.  Only recently was
the creation of a slowly moving
 pseudospin soliton investigated,\cite{kyr} but again this was done
near the balanced state.

 In this communication we study the dynamical problem of a strongly
 unbalanced double layer system in the presence of small or zero tunneling.
 Under such conditions the system is in highly nonequilibrium state but
 it is hard to reach the ground state if
the interlayer tunneling amplitude is small,
or even impossible if the amplitude is zero.
On the other hand
 the nonlinear coupling induced by interlayer coherence is still present
 and thus the appearance of different exotic nonlinear
 creations might be
expected.  Indeed, as we will show below, stable, mobile or spatially
 localized pseudospin textures arise which could be described in terms
 of ac driven nonlinear Schr\"odinger (NLS) equation widely studied in
 literature (see e.g.  Refs.  \cite{baras,baras2}).  This circumstance
 relates quantum Hall (pseudo)ferromagnets with completely different
 physical systems, such as yttrium iron garnet thin magnetic films
 \cite{kalinikos}, long Josephson junctions \cite{lomdahl}, relativistic
 electron-positron plasma \cite{berezhiani}, etc., which are governed by
 the same NLS equation and where similar localized excitations are well
 studied both experimentally and theoretically.

 The phenomelogical model for double layer quantum Hall (pseudo) ferromagnets,
 which is based upon microscopic considerations \cite{moon,yang},
 is effectively described by the following Hamiltonian
 \cite{moon,yang,brey,hanna}:
 \begin{eqnarray}
 {\cal H}&=&-\frac{eV}{2}n_z+\frac{\rho_E  }{2}\left\{\left(
 \frac{\partial n_x}{\partial x}\right)^2
 +\left(\frac{\partial n_y}{\partial x}\right)^2\right\}
 +\beta \,\,n_z^2
 \nonumber\\&&
 -\Delta_{SAS}\Bigl\{n_x{\cos}(Qx)+n_y{\sin}(Qx)\Bigr\},
 \label{1}
 \end{eqnarray}
 where $\rho_E $ is in-plane spin stiffness; $\beta$ is the a hard axis
 anisotropy;
 $\Delta_{SAS}$ is a tunneling amplitude, $Q\sim B_\parallel$ is the
 ``wave vector" associated with the tumbling of the electronic phase due to
 the in-plane magnetic field; $V$ describes the gate
 voltage; $e$ is an electron charge; $\hbar=1$ and $\hat n(x,t)$ is an
 order parameter unit vector. The component $n_z(x,t)$ is the
 local charge imbalance between the layers and can be expressed
 in terms of local filling factors of top and bottom layers,
 $n_z=\nu_1-\nu_2$; the local filling factors
$\nu_1$ and $\nu_2$ are proportional to the local
 electric densities $N_1$ and $N_2$
 in the corresponding layers. For example, in the fully balanced
 case, $\nu_1=\nu_2=1/2$ (i.e. $n_z=0$) and $N_1=N_2\equiv N$. In typical
 experiments \cite{spielman}
 on double layer systems $N=3.0\times 10^{10}cm^{-2}$. In the strongly
 unbalanced situation which is a subject of study in the present paper,
 one has $n_z\simeq 1$ and thus the densities of top and bottom
 layers are close to $2N$ and $0$, respectively.

 The time-space behavior of ordering vector could be described in terms of
 Landau-Lifshitz equation\cite{slichter}:
 \begin{equation}
 \frac{\partial {\hat n}}{\partial t}=\bigl({\hat n}\times{\vec
 H}_{eff}\bigr), \label{4}
 \end{equation}
 $$
 {\vec H}_{eff}=-2\left\{\frac{\partial{\cal H}}{\partial{\hat
 n}}-\frac{\partial}{\partial x}\left[\frac{\partial{\cal H}}{\partial
 \frac{\partial\hat n}{\partial x}}\right]\right\},
 $$
 where $\vec H_{eff}$ is the effective (pseudo)magnetic field.
 We introduce the variables
 \begin{equation}
 n^\pm\equiv n_x\pm i\, n_y
 \end{equation}
 so that eq.~(\ref{4}) has the following form:
 \begin{eqnarray}
 \frac{\partial n^+}{\partial t}&=&-ieVn^++4i\beta n^zn^++2i\rho_E
 n^z\frac{\partial^2n^+}{\partial x^2}
 \nonumber\\&&
 +2i\Delta_{SAS}n_ze^{iQx}. \label{pop}
 \end{eqnarray}
 We will seek for the solution of this equation around $n_z=1$. This state
 corresponds to the highest energy value static solution. Any dynamical
 fluctuation around this state is the precession around $n_z$ which
 is allowed by the equations of motion. In the limit of
 small deviations from this state and small tunneling it is possible to apply the
 multiple scale approach \cite{taniuti,ramaz}:
 $$
 n^+=\varepsilon m^+(\xi,\tau)e^{i(Qx-\omega t)},
$$
\begin{equation}
  n_z=\sqrt{1-|n^+|^2}=1-
 \varepsilon^2\frac{|m^+|^2}{2}, \label{def}
\end{equation}
$$
 \Delta_{SAS}\equiv\varepsilon^3 \tilde\Delta_{SAS}
$$
 where $m^\pm(\xi,\tau)$ is a slowly varying functions of the variables
 \begin{equation}
 \xi=\varepsilon(x-vt); \qquad \tau=\varepsilon^2 t \label{variable}
 \end{equation}
 and $\varepsilon$ is a formal small parameter expressing the smallness
 or ``slowness"
 of the object before which it appears, allowing for a multiple scale
 analysis of the problem.
The range of acceptable $\varepsilon$ will be determined in the
course of the calculation.
We are working in the regime:
 \begin{equation}
 |n^+|=|m^+|\ll 1. \label{cond}
 \end{equation}
 To study the perturbative solution we substitute eq.(\ref{def}) into the
 eq.(\ref{pop}), and collect terms of the same order of
 $\varepsilon$.  In the first order of
 $\varepsilon$ the following equality is obtained:
 \begin{equation}
 \omega=eV+2(\rho_E  Q^2-2\beta), \label{w}
 \end{equation}
 while to second order we get the expression for the group velocity $v$:
 \begin{equation}
 v=4\rho_E  Q. \label{v}
 \end{equation}
 Finally, to third order in $\varepsilon$ we come to the following nonlinear
 equation:
 \begin{eqnarray}
 i\frac{\partial m^+}{\partial \tau}+2\rho_E
 \frac{\partial^2m^+}{\partial\xi^2}+
 (\rho_E  Q^2-2\beta)|m^+|^2m^+=
 \nonumber\\
 2\tilde\Delta_{SAS} \,\,e^{i\omega t}, \hspace{2cm} \label{ac}
 \end{eqnarray}
 which in the case of
 \begin{equation}
 \rho_EQ^2>2\beta; \qquad
 \frac{eV+2\rho_EQ^2-4\beta}{\rho_EQ^2-2\beta}\sim\varepsilon^2 \label{rest}
 \end{equation}
 reduces to the exactly solvable ac driven NLS equation considered in Refs.
 \cite{baras,baras2}.

 The physical meaning of the second expression from (\ref{rest})
 together with definition (\ref{w}) is that one has a near resonance
 driving of pseudospin mode with a wavenumber $Q$.  The possibility of
 soliton, chaotic or other exotic solutions of the NLS
 equation (\ref{ac}) is studied in details in Refs.~\cite{baras,baras2}
 and we direct the reader there for more information.  Here we only
 mention that the stable solutions of equation (\ref{ac})
 could be interpreted as envelope solitons of order parameter $\hat n$
 moving with a group velocity proportional to applied in-plane magnetic
 field [see expression (\ref{v})] and characterized by carrier wave
 vector $Q$ which is also proportional to in-plane magnetic field.
 Note that according to the restriction of eq.(\ref{rest}), the in-plane
 magnetic field should be nonzero.

\begin{figure}[htp] 
page \begin{center}\leavevmode
\includegraphics[width=0.85\linewidth]{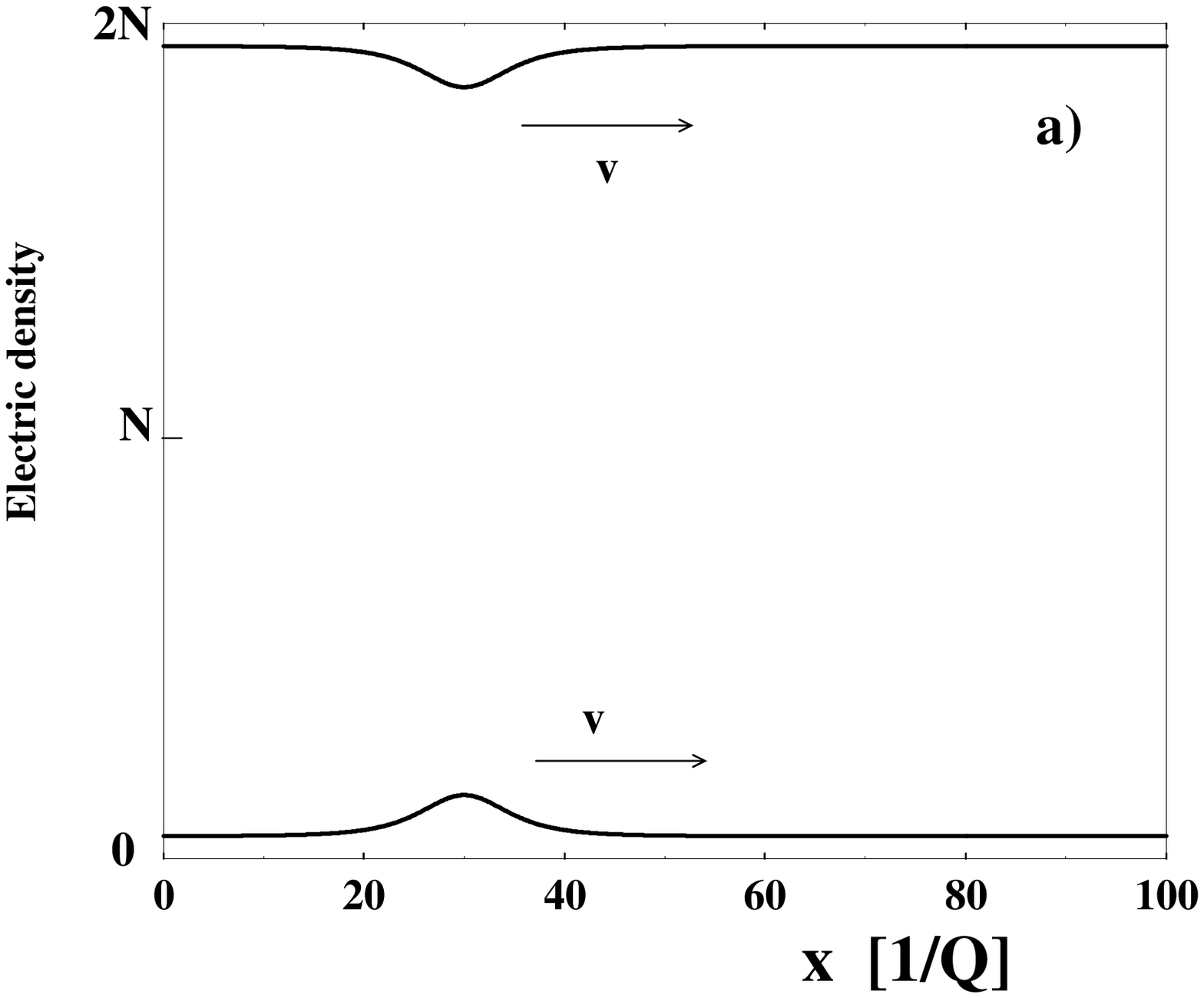}
\label{mama.eps}\end{center}\end{figure}

\begin{figure}[htp]
\begin{center}\leavevmode
\hspace{0.8cm}
\includegraphics[width=0.7\linewidth]{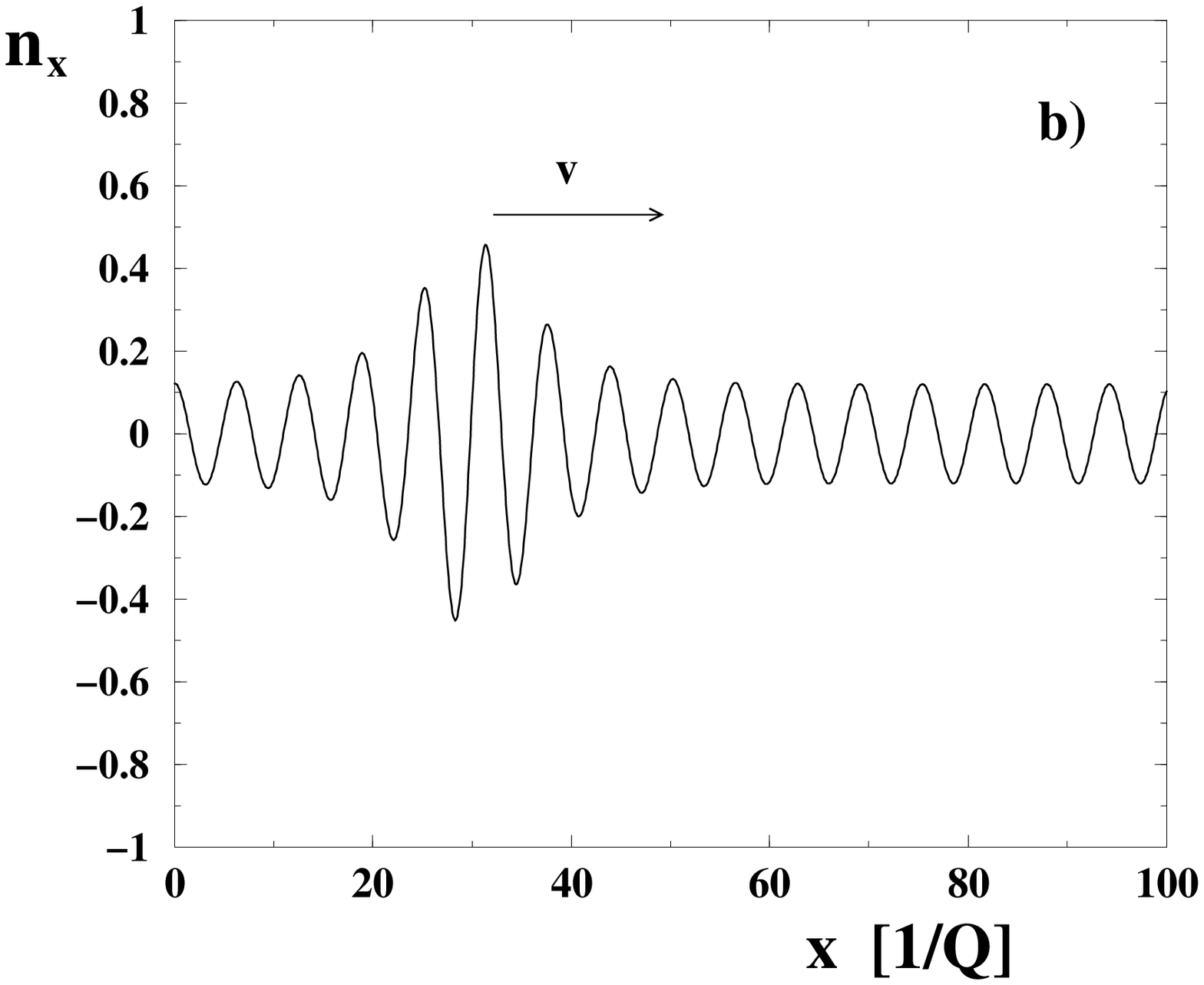}
\vspace{1.5cm}
\caption{(a) Local electric density of each layer versus $x$ in the case of small
tunneling. Top and bottom graphs
correspond to the local electric densities of top and bottom layer, respectively.
$N$ is an electric density of each layer in the fully balanced state.
(b) One of the transverse components of ordering vector (the second one is shifted
in phase by 90 degree) with carrier wave number $Q$. This envelope soliton
moves with velocity $v$ proportional
to the applied in-plane magnetic field [see also Exp. (\ref{v})]}
\label{mamb.eps}\end{center}\end{figure}

 In Fig. 1 this excitation is presented
 as a moving bump of electric density through each layer. Note that envelope
 soliton nature only characterizes the transverse component of order parameter,
 while the physically measurable quantity - local charge imbalance -
is just an ordinary soliton.

  According to the general results of Refs. \cite{baras,baras2}
 the stability of the mentioned soliton is restricted to the following condition:
 \begin{equation}
 h\equiv\frac{\Delta_{SAS}\sqrt{\rho_EQ^2-2\beta}}{(eV+2\rho_EQ^2-4\beta)^{3/2}}
 <\sqrt{\frac{1}{27}}.
 \end{equation}
 This quantity stands for the driving force in NLS equation \cite{baras}
 and therefore defines the soliton amplitude.

 Here we also present the solutions for simple,
 but physically very important case
 of zero tunneling ($\Delta_{SAS}=0$),
which do not directly follow from the exact
 solutions for nonzero $\Delta_{SAS}$ presented in Ref.~\cite{baras}.
 In the case of zero tunneling the in-plane magnetic
 field has no effect on the system.
 For simplicity we assume also $V=0$. Then the solution of the initial
 equation of motion (\ref{pop}) is sought for in the following simple form:
 \begin{equation}
 n^+=\varepsilon m^+(x,t)e^{4i\beta t}, \qquad
 n_z=1-\varepsilon^2\frac{|m^+|^2}{2}.
 \label{def1}
 \end{equation}
 Substituting this expression into Eq. (\ref{pop}) we come to the ordinary
 NLS equation:
 \begin{equation}
 i\frac{\partial m^+}{\partial t}+2\rho_E  \frac{\partial^2m^+}{\partial
 x^2}-2\beta
 |m^+|^2m^+=0, \label{nac}
 \end{equation}
 which has a stable ``dark soliton'' solution \cite{abl}.
 Since the physically measurable variable is
 $n_z$ (the local charge imbalance between the layers)
 we present here only an expression for its profile:
 \begin{equation}
 n_z=1-\frac{D^2}{2}\left|\frac{\sqrt{1-A^2}}{A}+i\cdot
 \tanh\left\{\frac{x}{\Lambda}\right\}\right|^2,
 \label{13a}
 \end{equation}
 where $D$ is a soliton amplitude; $A$ denotes the contrast of dark soliton
 (if $A=1$ one has a ``black'' dark soliton and ``gray''
 dark for $0<A<1$) and soliton width $\Lambda$ is defined as follows:
 \begin{equation}
 \Lambda=\sqrt{\frac{2\rho_E}{\beta}}\frac{1}{D}. \label{13b}
 \end{equation}

 The difference between gray dark and black dark solitons
can be expressed the
in terms of the
 charge densities in the top and bottom layers
 (see also Fig.  2):  In a black dark
 soliton solution there exists some point along $x$ axis where the local
 electric density of top layer reaches the value $2N$, while the local
 density of bottom layer in the same point approaches zero.  On the
 other hand, there always exists some deviation of the  charge densities
 from $2N$ and $0$ for top and bottom layers, respectively, for the
 gray dark soliton solution.

 Unlike the previous case of small tunneling where the soliton is
 characterized by nonzero carrier wave number $Q$, in the case of zero
 tunneling that is not the case and consequently transverse components
 of ordering vector have the similar form to the $z$ component and are
 not presented in Fig.  2.

 \begin{figure}[htp] 
\begin{center}\leavevmode \includegraphics[width=0.9\linewidth]{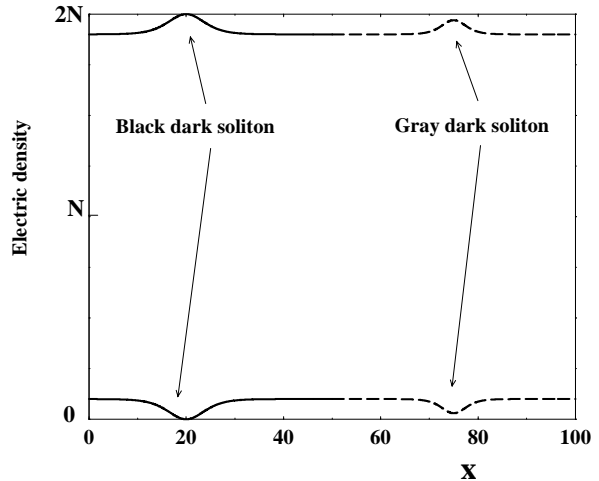}
\caption{Static ``bumps" of electric density for zero tunneling in
strongly unbalanced double layer system.  Two types of solutions are
presented:  solid line corresponds to black dark soliton solution, while
dashed line indicates gray dark soliton.  Both soliton solutions are
stable.}  \label{mam.eps}\end{center}\end{figure}

 We should emphasize again one more difference between two considered
 cases:  In the case of nonzero (but small) tunneling the envelope soliton
 amplitude is linked with the driving coefficient $\Delta_{SAS}$, while in
 the absence of tunneling the dark soliton amplitude is arbitrary
 within the restriction (\ref{cond}).

 The above considerations are only valid for quasi-one dimensional
 samples, i.e.  quantum Hall bars.  Thus the main requirement for
 stability of soliton solutions is that the transverse dimension of the
 double layer system should be less than soliton width.  In this case
 the large wavelength modulations do not grow (which destabilize the
 soliton) and solitons remain stable \cite{zakharov,kivshar}.  Future work
will investigate their stability.

 In conclusion, we have found stable (mobile or static)
 charge-density excitations in strongly unbalanced double
 layer systems.  For their observation the double layer quantum Hall
 system should be prepared so as $n_z\simeq 1$.  Then any perturbation
 will cause the appearence of either standing (zero tunneling) or moving
 (for small tunneling and applied in-plane magnetic field) charge density
 train of ``bumps", which could be easily detected experimentally.  The
 perturbation could be induced by temporary application of local electric
 field perpedicular to the layers.

 The work at the University of Oklahoma was supported by the NSF under grant No.
 EPS-9720651 and a grant from the Oklahoma State Regents for higher education.
 R. Kh. was supported by the
 NSF-NATO visiting scientist fellowship award No DGE-0075191.
R.A. was supported by NSF MRSEC Grant DMR-0080054.

\end{document}